\documentclass[aoas,preprint]{imsart}
 
\usepackage{amsthm,amsmath,natbib}
\RequirePackage[colorlinks,citecolor=blue,urlcolor=blue]{hyperref}
\usepackage{mathptmx}       
\usepackage{helvet}         
\usepackage{courier}        
\usepackage{type1cm}        
\usepackage{amsfonts}
\usepackage{amsbsy}
\usepackage{comment}
\usepackage{makeidx}         
\usepackage{graphicx}        
\usepackage{dsfont}                             
\usepackage{multicol}        
\usepackage{multirow}
\usepackage[bottom]{footmisc}%
\usepackage{amssymb}
\usepackage{color} 
\usepackage[export]{adjustbox}
\usepackage{enumitem}
\usepackage[font=small,labelfont=bf,tableposition=top]{caption}
\DeclareCaptionLabelFormat{andtable}{#1~#2  \&  \tablename~\thetable}

\DeclareMathOperator{\D}{\mathcal D} 
\DeclareMathOperator{\sD}{\tilde{\mathcal D}} 
\DeclareMathOperator{\A}{\mathcal A} 
\DeclareMathOperator{\E}{\mathbb E}

\DeclareMathOperator*{\argmax}{arg\,max} 
\DeclareMathOperator{\f}{\pmb f} 
\DeclareMathOperator{\s}{\pmb s} 
\DeclareMathOperator{\p}{\mathbb{P}} 
\DeclareMathOperator{\x}{\pmb x} 
\DeclareMathOperator{\z}{\pmb z}
\DeclareMathOperator{\y}{y} 
 
\DeclareMathOperator{\Cov}{\mathrm{Cov}} 
 
\newtheorem{Lemma}{Lemma}
\startlocaldefs
\endlocaldefs

\begin{document}

\begin{frontmatter}

\title{Bayesian model-based spatiotemporal survey design for log-Gaussian Cox process}
\runtitle{Survey design for log-Gaussian Cox process}



\begin{aug}
\author{\fnms{Jia} \snm{Liu}\thanksref{t2}\ead[label=e1]{jia.liu@jhelsinki.fi}},
\author{\fnms{Jarno} \snm{Vanhatalo}\thanksref{t2,t3}\ead[label=e2]{jarno.vanhatalo@helsinki.fi}}

\thankstext{t2}{Department of Mathematics and Statistics, Faculty of Science}
\thankstext{t3}{Organismal and Evolutionary Biology Research Program, Faculty of Bio- and Environmental Sciences}
\runauthor{J. Liu and J. Vanhatalo}

\affiliation{University of Helsinki }

\address{Department of Mathematics and Statistics,\\
Faculty of Science,  University of Helsinki,\\
P.O. Box 68, FI00014 Finland \\
\printead{e1}\\
\phantom{E-mail:\ jia.liu@jhelsinki.fi \ }}

\address{Department of Mathematics and Statistics, \\
Faculty of Science, University of Helsinki, \\
 P.O. Box 68, FI00014 Finland \\
Organismal and Evolutionary Biology \\ Research Program,  
Faculty of Bio- \\ and Environmental Sciences, \\
 University of Helsinki, \\
 P.O. Box 68, FI00014 Finland\\
\printead{e2}\\
\phantom{E-mail:\ jarno.vanhatalo@helsinki.fi\ }} 
\end{aug}

\runauthor{J. Liu and J. Vanhatalo }

\begin{abstract}
In geostatistics, the design for data collection is central for accurate prediction and parameter inference.  
One important class of geostatistical models is log-Gaussian Cox process (LGCP) which is used extensively, for example, in ecology. 
However, there are no formal analyses on optimal designs for LGCP models.
In this work, we develop a novel model-based experimental design for LGCP modeling of spatiotem\-po\-ral point process data. 
We propose a new spatially balanced {\it rejection sampling design} which directs sampling to spatiotemporal locations that are \emph{a priori} expected to provide most information.
We compare the rejection sampling design to traditional balanced and uniform random designs using the average predictive variance loss function and the Kullback-Leibler divergence between prior and posterior for the LGCP intensity function.
Our results show that the rejection sampling method outperforms the corresponding balanced and uniform random sampling designs for LGCP whereas the latter work better for models with Gaussian models. 
We perform a case study applying our new sampling design to plan a survey for species distribution modeling on larval areas of two commercially important fish stocks on Finnish coastal areas. 
The case study results show that rejection sampling designs give considerable benefit compared to traditional designs. 
Results show also that best performing designs may vary considerably between target species.
\end{abstract}


\begin{keyword}
\kwd{Experimental design; survey design; Bayesian inference; Kullback-Leibler information; log Gaussian Cox process; spatiotemporal; ecology; species distribution}
\end{keyword}

\end{frontmatter}

\section{Introduction}

A central question of geostatistics is the prediction of a spatial patterns over a region using data measured at a finite set of locations. 
A widely used stochastic model for such tasks is a hierarchical Gaussian process model
\citep{Cressie:1993,gelfand2010}. 
It is well known that the goodness of the spatial prediction with such models depends on the spatial allocation of the measurement locations \citep{muller2001,diggle2006} -- that is on the \emph{observational/ex\-pe\-ri\-men\-tal design}. 
The problem of finding an optimal design for spatial prediction when the observation process is Gaussian has received much interest in spatial statistics \citep[e.g.,][]{muller1999,muller2001,diggle2006}. What has been left for lesser attention are spatiotemporal designs and designs for models with non-Gaussian observation processes \citep[see][for few examples]{chipeta2015,chipeta2016}. 
In this work, we study observational designs for spatiotemporal log-Gaussian Cox processes (LGCPs). 

In traditional application of LGCPs, the spatial study region is observed fully and the statistical analysis reduces to inference concerning the underlying intensity function ~~~~~~\citep{Moller_etal:1998,Moller_etal:2007}. Consider, for example, the classical analysis of locations of trees in a forest \citep{illian+etal:2012,Simpson2015a}. However, in recent years LGCPs have gained increasing interest in applications where the study region is not fully observed. For example in ecology, LGCPs are used in species distribution modeling \citep{Warton2010,Chakraborty2011,Renner2013,Yuan2016} where observations comprise of animal counts or presence observations at, for example, survey plots or transects that cover only small proportion of the whole study region. In these applications the LGCP describes the process generating locations of individual animals or plants and the observations are a thinned version of the underlying LGCP. The thinning process describes the (relative) probability of observing individual points (i.e., animals or plants) and it can vary from zero to one within the study region, depending on the sampling effort. The statistical analysis includes predicting the function also in places where observations have not been made \citep[e.g.][]{Yuan2016,Kallasvuo+etal:2017,Vanhatalo+Hosack+Sweatman:2017} resulting in a spatial prediction problem where sampling design plays critical role.

Model-based optimal experimental design concerns a problem of maximizing the expected utility of future data over alternative designs.
Alternatively, we may consider minimizing the expected loss.
Much of the literature \citep[e.g.][]{stein1989,zhu2006} is dedicated to developing appropriate utility functions and computational methods. 
In most cases, the expectation is analytically intractable and optimization over the design space computationally demanding.
Common methods to approximate these include Monte Carlo simulation \citep{robert2004,vlachos1996,chipeta2015} or a series of simulated annealing algorithms \citep{van1998,muller1999}. 
Alternatively, many authors such as \citet{muller2001}, \citet{muller2007}, \citet{ryan2015} and \citet{chipeta2016} have aimed at developing spatially balanced designs that increase expected utility for a given class of spatial models over that of uniform random designs. 
In these approaches, the computation of the utility is a minor task since the expected utility of the candidate designs has to calculated only once.  

Spatially balanced designs provide more uniform coverage over the study area than random sampling \citep{robertson2013b}, which decreases uncertainty in spatial interpolation.
A balanced design maintains spatial regularity of sampling locations by spreading observation points as evenly as possible in the design space by means of specific sampling methods or criteria \citep{muller2007,stevens2004,grafstrom2012}.
Quasi-random methods use quasi-random number generators such as the Sobol and Halton sequence to generate balanced and well-distributed designs. 
Distance-based designs are deterministic algorithms that aim to produce space filling designs by restricting the minimum distance between any two sampling locations \citep{chipeta2016,russo1984,royle1998}.
Sampling locations in the above mentioned designs are commonly selected with equal probability. 
This may be inappropriate in applications where some locations are \emph{a priori} expected to be more informative than others.
As we demonstrate here, this happens with LGCPs when there is prior knowledge at its intensity function. 
Such prior information may arise, for example, from earlier studies and can be used to more efficiently plan new studies.

In this work, we are specifically interested in spatiotemporal design problems and propose to extend common spatially-balanced designs with a rejection sampling scheme that gives more weight to times and subregions which are \emph{a priori} expected to be more informative about the intensity function of the point process. 
A novel sampling design method is developed that enables prior knowledge to be imposed from any or all covariates in the survey design. 

The structure of this paper is as the follows. In section~\ref{sec:model} we describe LGCP modeling and present motivating examples. 
We then revise the average predictive variance loss function and the Kullback-Leibler (KL) divergence utility function as two common criteria for design evaluation (Section~\ref{sec:utilities}) and discuss some of their properties in LGCPs (Section~\ref{sec:comp+properties}). 
Next we review the widely used balanced designs, and propose a novel rejection sampling design method (Section~\ref{sec:design}). 
In section~\ref{sec:simulation_studies}, we examine the designs with simulation studies. A case study of survey design concerning fish reproduction areas in fisheries management is presented in Section~\ref{sec:case_study}. We end with discussion and conclusions in Section~\ref{sec:discussion}

\section{Log Gaussian Cox processes}\label{sec:model}

\subsection{The model and motivating example}
The main motivation for our work comes from species distribution modeling where LGCPs have received increasing interest in recent years \citep[e.g.,][]{Warton2010,Chakraborty2011,Renner2013,Yuan2016,Makinen+Vanhatalo:2018}. 
We use species distribution modeling as a running example in this work even though the model arises in other applications as well. 

We denote the study domain of interest by $\D \subset \A\times[t_o,t_1]$ where $\A\subset \Re^2$ is the spatial region and $[t_o,t_1]$ the time interval of interest. 
A spatiotemporal LGCP arises from an inhomogeneous Poisson process with a spatially and temporally varying intensity whose logarithm is given a Gaussian process prior \citep{Chakraborty2011,Banerjee+Carlin+Gelfand:2015}. 
We denote the intensity function by $\lambda(\s,t) = \lambda(\x)$ where $\x=[\s^{T},t]\in \D$ is a vector of spatiotemporal coordinates.
The Gaussian process prior for the logarithm of the intensity means that any finite collection of latent variables $f_i = \log \lambda(\x_i), i=1,...,n$ has a multivariate Gaussian distribution. 
A Gaussian process is defined by its mean, $\mu(\x)=\E\left(f(\x)\right)$, and covariance, $k\left(\x,\x';\theta\right)=\Cov\left(f(\x),f(\x')\right)$, functions where $\theta$ are the hyperparameters of the covariance function. 

In species distribution modeling, $\lambda(\x)$ corresponds to the (relative) density of a species at location $\s$ at time $t$. One typically does not observe individual animals or plants with probability one and only seldom is the whole study domain surveyed. Hence, the observations are a realization of a thinned LGCP with intensity function $\lambda(\x)\pi(\x)$ where $\pi(\x)$ is the observation probability. The observation probability can be anything from zero to one but for simplicity we assume that it is either zero or one here. This corresponds to a situation where the study domain is surveyed only partially \citep{illian+etal:2012,Simpson2015a}.

Since the likelihood function of a LGCP includes integral over the domain, the domain is commonly discretized so that the likelihood can be approximated by the product of finite number of Poisson likelihood terms, one for each discretized location \citep{Banerjee+Carlin+Gelfand:2015}. If the domain of interest is not fully observed, the (approximate) likelihood can be written so that only the discretized locations within observed areas contribute to it \citep[see, e.g.,][]{Chakraborty2011,Simpson2015a}.
That is, the likelihood function is
\begin{align}
L\left(y_1,\hdots,y_n|f(\cdot)\right) &= L(y_1,\hdots,y_n|\f)\nonumber \\
& =\prod_{i=1}^n \mathrm{Poisson}\left(y_i|\lambda(\x_i) \right)\label{eq:likelihood}
\end{align}
where $n$ is the number of observed discretized locations, $y_i$ is the count observation at $i$'th location $\x_i$, and $\f=[f(\x_1),\dots,f(\x_n)]^T$ is the vector of latent variables at locations. The locations $\x_i, i=1,\dots,n$ might correspond to, for example, surveyed lattice grid cells \citep{Chakraborty2011,Makinen+Vanhatalo:2018} or survey sites \citep{Vanhatalo+Hosack+Sweatman:2017}. 

As an example, \citet{Vanhatalo+Hosack+Sweatman:2017} modeled the spatiotemporal fluctuations of crown-of-thorns starfish in the Great Barrier Reef in the north-east Australia and
\citet{Kallasvuo+etal:2017} modeled fish larval areas along the Finnish coastline with the objective of finding spatial areas that produce most recruits to fish stocks. 
From a survey design point of view, the key question in these applications is where and when we should sample to learn most about the essential aspects of starfish density on reefs and fish larvae density in the water column. 
For both of these examples the existing data provide prior knowledge concerning the intensity which can be used in the planning of future surveys. 
A distinctive feature of the problem compared to traditional Gaussian process modeling with a Gaussian observation model is the expected utility of a design that is highly dependent on the prior knowledge of the mean intensity.
We consider two types of spatiotemporal Gaussian process priors for the log intensity
\begin{align}
\text{Separable model:\hspace{1cm}} & \log \lambda(\x)=f(\s,t) \sim GP(\mu(\s,t), k\left(\s,\s')k(t,t')\right) \label{eq:separable} \\ 
\text{Additive model:\hspace{1cm}} & \log \lambda(\x)=f(\s,t) \sim GP(\mu(\s,t), k\left(\s,\s')+k(t,t')\right). \label{eq:add}
\end{align}

The rationale for considering these two models is the following. 
The separable model is a commonly used ``general purpose'' spatiotemporal model whose covariance structure allows joint effects of space and time \citep{schmidt2003,kyriakidis1999}. 
The additive model corresponds to $f(\s,t)=\mu(\s,t) + g(\s)+h(t)$ where the additive terms are mutually independent Gaussian processes $g(\s)\sim GP(0,k(\s,\s')$ and $h(t)\sim GP(0,k(t,t'))$.
In the additive model, the spatial pattern of intensity is stable in time but there are spatially constant relative changes in intensity. 
This can be used to represent, for example, distributions of species that are present in their stable distribution area only at certain time of the year (see Section~\ref{sec:case_study}). 
In this case, the component $g(\s)$ has the interpretation of distribution area and $h(t)$ explains the temporal changes in their abundance. 
In both models, the mean, $\mu(\x)$, is a deterministic function that represents prior information about expected temporal changes in species intensity across the study domain.

\section{Design criteria} \label{sec:utilities}

\subsection{Expected utility and loss of a design}

We denote by $D_n=\{d_n\}$ the set of all possible designs $d_n$ of size $n$ in domain $\D$. 
For simplicity we assume that a design $d_n=\left\{\x_1,\dots,\x_n: \x_i \in \D\right\}$ is a collection of $n$ sampling sites with coordinates $\x_i$ corresponding to the discretized locations in \eqref{eq:likelihood}. That is, survey sites are discretized locations where $\pi(\x)=1$.
In general, utility of future data collection depends on design, future data and model parameters. 
We denote by $U(d_n,Y,f(\cdot),\theta)$ a utility function where $Y=[Y_1,\dots,Y_n]^T$ is a random vector denoting the new data to be collected at survey sites.
Alternatively, we may define a loss function $L(d_n,Y,f(\cdot),\theta)$.
In a more general treatment where surveys are used to inform decision making, utility and loss should depend also on decisions \citep{Lindley:2003,Eidsvik:2015}.
 However, we do not consider decision making here and omit decisions from our notation.
A design should be evaluated according to its expected utility which in the case of non-negative integer observations is
\begin{align}\label{eq:expected_utility}
\bar{U}(d_n) &= \sum_{y\in \mathbb{N}^n} p(y|d_n) \int_{f}\int_{\theta} U(d_n,y,f(\cdot),\theta)d \p(f(\cdot)|d_n,y,\theta)d\p(\theta|y,d_n) ,  
\end{align}
where $\p(f(\cdot)|d_n,y,\theta)$ and $\p(\theta|y,d_n)$ are the posterior probability measures of the latent function and the parameters given a realization $y=[y_1,\dots,y_n]$ from the design $d_n$ and $p(y|d_n)=\int p(y|f(\x_1),\dots,f(\x_n))dp(f(\x_1),\dots,f(\x_n))$ is the (prior) predictive density of $y$. Hence, the outer summation corresponds to expectation over the prior predictive distribution of $Y$.
In this work, we consider $\theta$ to be fixed so that the evaluation criteria is the expected utility conditional on $\theta$
\begin{align}\label{eq:expected_utility_fixed_theta}
\bar{U}(d_n) = \bar{U}(d_n,\theta= \tilde \theta) &= \sum_{y\in \mathbb{N}^n}  p(y|d_n)\int U(d_n,y,f(\cdot))d \p(f(\cdot)|d_n,y) .  
\end{align}
In simulation studies, we evaluate designs with alternative values of $\theta$ in order to gain understanding about the effect of hyperparameters on the expected utility or loss. 
In the case study, we fix $\theta$ to its \emph{maximum a posteriori} (MAP) estimate in order to make the computations feasible.

\subsection{Average predictive variance (APV) loss}
Spatial designs are commonly compared with the average predictive variance (APV) loss over the study domain. 
Here we take APV as our first criterion due to its popularity and its capability of informing the expected marginal accuracy of point wise predictions over the study domain \citep[see, e.g.,][]{diggle2006,chipeta2016}. 
The APV loss function depends only on the predictive variance of the latent function $f(\cdot)$ so that the loss function and the corresponding expected loss are given by 
\begin{align}\label{eq:u21}
L_{\text{APV}}(d_n,Y) &= \frac{1}{|\D|}\int_{\x_{\ast}\in\D} \mbox{Var}\{f(\x_{\ast})|d_n, Y\} d\x_{\ast},\\
\bar{L}_{\text{APV}}(d_n) &= \frac{1}{|\D|} \sum_{y\in \mathbb{N}^n} p(y|d_n) \int_{\x_{\ast}\in\D} \mbox{Var}\{f(\x_{\ast})|d_n, y\} d\x_{\ast} \label{eq:APVloss}.
\end{align}
%
%
where $|\D|$ is the size (area/volume) of the study domain. 
Similarly, we consider the expected APV loss of the intensity 
\begin{align}\label{eq:u2}
\bar{L}_{\text{APV}\lambda}(d_n) &= \frac{1}{|\D|} \sum_{y\in \mathbb{N}^n} p(y|d_n) \int_{\x_{\ast}\in\D} \mbox{Var}\{\lambda(\x_{\ast})|d_n, y\} d\x_{\ast}.
\end{align}
With a log transformation, $\lambda(\x_{\ast})=\exp(f(\x_{\ast}))$, the mean and variance of $\lambda(\x_\ast)$ are 
\begin{align*} 
\mu( \lambda(\x_\ast)) &= \displaystyle \exp\biggl(\mu(f(\x_\ast)) +\mbox{Var}(f(\x_\ast))/2\biggr),  \\
\mbox{Var}[ \lambda(\x_\ast)] &= \displaystyle \biggl[\exp(\mbox{Var}(f(\x_\ast)) - 1 \biggr]\exp\biggl(2\mu(f(\x_\ast)) +\mbox{Var}(f(\x_\ast))\biggr).
\end{align*}
We approximate the integral over $\D$ by discretizing the study domain into lattice grids $X_{\ast}=\{\x_{\ast,1},\dots,\x_{\ast,N}\}$ with $N$ cells and coordinates $\x_{\ast}$ over which we take a finite sum. 
The expectation over $Y$ is approximated by Monte Carlo approximation 
\begin{align*}
\bar{L}_{\text{APV}}(d_n)  \approx \frac{1}{M} \sum_{j=1}^M \biggl[ \frac{1}{N} \displaystyle \sum_{\x_{\ast}\in X_{\ast}} \mbox{Var}\{f_j(\x_{\ast})|d_n, Y_j\}\biggr], 
\end{align*}
where $Y_j \in \mathbb{N}^n $ is the $j$'th Monte Carlo draw from $\p(Y|d_n)$ and $M$ is the number of Monte Carlo samples.

\subsection{The expected Kullback-Leibler divergence}

In many cases, for example when estimating integrals or sums over spatial regions, point wise predictions are not enough but we need to know the full posterior \citep[see, e.g,][]{Kallasvuo+etal:2017}. 
Then the aim of data collection is to increase the information concerning the joint distribution of the latent and intensity function. 
In this case, a natural choice for the utility function is the Kullback-Leibler (KL) divergence from prior to posterior \citep{Lindley1956,kullback1987} which for a LGCP is
\begin{align}\label{eq:ekl1}
 U_{\text{KL}}(d_n,Y) &= \mbox{KL}\biggl(d \p(f(\cdot)|X,Y)||d \p(f(\cdot)\biggr)  \\
\bar{U}_{\text{KL}}(d_n) &= \sum_{y\in \mathbb{N}^n} p(y|d_n) \mbox{KL}\biggl(d \p(f(\cdot)|X,y)||d \p(f(\cdot))\biggr) \label{eq:KL2}.
\end{align}
We use again Monte Carlo approximation to numerically solve the expectation over $Y$ in \eqref{eq:KL2}. 
The KL-divergence has a particularly simple form when the observations \\ $Y_1,\dots,Y_n$ are conditionally independent given the corresponding latent variables:
\begin{Lemma}\label{le:1}
Assume $f(\cdot)$ is a latent function with Gaussian process prior probability measure $\p(f(\cdot))$. 
Assume further that we have finite data $(d_n,Y)$, where $d_n=[\x_1^T,\dots,\x_n^T]$ are covariates and $Y=[Y_1,\dots,Y_n]$ are observations that are conditionally independent given the corresponding latent variables, that is $p(Y|f(\cdot))= \prod_{i=1}^n p(Y_i|f(\x_i))$. 
Denote by $\p(f(\cdot)| d_n,Y)$ the posterior probability measure of  $f(\cdot)$. 
The KL-divergence from the prior to the posterior for $f(\cdot)$ is
\begin{align}&\label{eq:KL}
\mbox{KL}\biggl(d \p(f(\cdot)|d_n,Y)||d \p(f(\cdot))\biggr)  &\notag\\&
=  \displaystyle \sum_{i=1}^n \int      p(f(\x_i)|d_n,Y) \log p(Y_i| f(\x_i) )  d f(\x_i)   - \log p(Y),
\end{align}
where $\log p(Y)=\log \int p(Y|\f)p(\f)d\f$ is the log marginal likelihood.

\end{Lemma}
See Appendix~\ref{apt:subc1} for proof.
Hence, in order to calculate the KL divergence from the prior process to the posterior process over $\D$ we need to calculate the marginal likelihood $p(Y)$ and $n$ one dimensional integrals. 
In case of a Gaussian observation model, these are analytically available and with LGCP we can use, for example, Laplace approximation (Section \ref{sec:laplace_approx}). 
The KL divergence in case of intensity $\lambda(\cdot)$ is the same as the KL divergence of the latent process $f(\cdot)$ (see Appendix~\ref{apt:C}).


Sometimes the interest is to predict the latent function and intensity only over a subdomain $\sD \subset \D$ which does not contain all observation locations (see Section~\ref{sec:case_study}).
Let's denote by $f_{\sD}$ the latent field over subdomain $\sD$. The KL-divergence from prior to posterior for $f_{\sD}$ is
\begin{align}&\label{eq:klfield}
\mbox{KL}\biggl(d \p(f_{\sD}|d_n,Y)||d \p(f_{\sD})\biggr)   
= \int \log p(y| f_{\sD}) d \p (f_{\sD}|_n,Y) - \log p_{\sD}(Y),
\end{align}
where  $p_{\sD}(y)= \int p(y| f_{\sD}) d\p(f_{\sD})$. Note that $p(y| f_{\sD})= \int p(y| \f) d \p(\f| f_{\sD})$. Hence, calculating the KL-divergence becomes more difficult than in \eqref{eq:KL} (Appendix~\ref{apt:subc1}).

\section{Computation and properties of the expected utility and loss}\label{sec:comp+properties}

\subsection{Approximate posterior inference with Laplace approximation}\label{sec:laplace_approx}
The traditional method to infer the LGCP is Markov chain Monte Carlo \citep{Moller+Waagepetersen:2004} but in recent years analytic approximations such as the Integrated Nested Laplace approximation \citep{illian+etal:2012,Simpson2015a} and Gaussian approximations built with expectation propagation and Laplace method \citep{Vanhatalo+Pietilainen+Vehtari:2010,Kallasvuo+etal:2017} have become popular due to their computational benefits. Here, we use the Laplace method built over the Gaussian approximation due to its simple analytical form and because it has been shown to give accurate approximation for these models \citep[e.g.,][]{Vanhatalo+Pietilainen+Vehtari:2010}.

For a given design $d_n$ and realization of observations $y$, Laplace approximation for the posterior of the latent function, conditional on $\theta$, is $p(f(\x_{\ast})|d_n, y, \theta) \approx N\left(f(\x_{\ast})|\mu_{\ast|y},K_{\ast|y}\right)$, where the (approximate) posterior mean and variance are 
\begin{align}
\mu_{\ast|y,\theta}& = \mu(\x_\ast)+K(\x_{\ast},d_n)K(d_n)^{-1}(\hat{\f}-\mu(d_n)) \label{eq:post_mean}\\
K_{\ast|y,\theta}&= \mbox{Var}\{f(\x_{\ast})\} - K(\x_{\ast},d)(K(d_n)+W_y^{-1})^{-1}K(d_n,\x_{\ast}) .\label{eq:post_var} 
\end{align}  
Here, $K(\x_{\ast},d_n)$ is the prior covariance matrix between the prediction location $\x_{\ast}$ and all the design locations; $K(d_n)$ and $\mu(d_n)$ are the prior covariance matrix and mean vector at the design locations; $\hat{\f} = \argmax_{\f} p(\f|y,d_n)$ is the \emph{maximum a posteriori} (MAP) estimate of latent variables; and $W_y=-\nabla\nabla \log L(y|\f)|_{\f=\hat{\f}}=\text{diag}(e^{\hat{f}_1},\dots,e^{\hat{f}_n})$ is the negative Hessian matrix of the log likelihood. 
The Laplace approximation for the joint posterior of $\f_{\ast}$ is formed similarly by replacing \\ $\mbox{Var}\{f(\x_{\ast})\}$ with the prior covariance matrix of $\f_{\ast}$, to be denoted by $K_{\ast}$.

\subsection{Properties of the utility and loss criteria}\label{sec:properties_of_utilities}

Recall the Laplace approximation for the conditional posterior mean and variance of the latent variables \eqref{eq:post_mean}-\eqref{eq:post_var}.
By Woodbury-Sherman-Morrison Lemma we can write $(K(d_n) + W_y^{-1})^{-1}=W_y-W_y(K(d_n)^{-1}+W_y)^{-1}W_y$. When $\theta$ are fixed, the posterior predictive variance is controlled by $W_y$, which decreases proportionally with $\hat{\f}$: $W_y\rightarrow 0$ as $\hat{\f}\rightarrow -\infty$ such that  the posterior variance reverts to the prior variance.
On the other hand, 
\begin{equation}
\hat{\f}|Y,d_n = \argmax_{\f} -(\f-\mu)^TK(d_n)^{-1}(\f-\mu) + \sum_{i=1}^n \left( -e^{f_i} + Y_if_i \right)
\end{equation}
where $\E[Y_i] = e^{\mu_i(d_n)+2K(d_n)_{ii} }\rightarrow 0$ when $\mu_i(d_n)\rightarrow -\infty$. For this reason,  $\E_{p(Y)}\{\hat{\f}\}\rightarrow \mu(d_n)$ when $\mu(d_n)\rightarrow -\infty$. 
Intuitively the posterior mean and covariance of $f(\cdot)$ are not \emph{a priori} expected to change from the prior covariance of $f(\cdot)$ if we sample from areas where $\mu(\x)$ is so small that the prior predictive probability $\mbox{Pr}(y_i>0)\approx 0$. 
The difference between prior and posterior should be larger the more design points are located in places where the prior predictive probability $\mbox{Pr}(y_i>0)$ is significantly above zero. 
This property is very different from the properties of a Gaussian process with Gaussian observation model $y_i|f(\x_i)\sim N(f(\x_i),\sigma^2)$ where the posterior predictive mean and variance are 
\begin{align}
\mu_{\ast|y} &= \mu_{\ast} + K(X_{\ast},d_n)(K(d_n)+ \sigma^2_n I)^{-1} (y-\mu(d_n)) \label{eq:Gaussian_postMean}\\
K_{\ast|y,\theta} &= \mbox{Var}\{f(\x_{\ast})\} - K(\x_{\ast},d_n)(K(d_n)+\sigma^2I)^{-1}K(d_n,\x_{\ast})\label{eq:Gaussian_postVar},
\end{align} 
Under a Gaussian observation model $\bar{L}_{\text{APV}}(d_n)$ does not depend on $\mu(\s,t)$ so a uniform space filling design works well.
Similarly for the KL-divergence utility, the locations where $\mbox{Pr}(y_i>0)\approx 0$ are \emph{a priori} expected to be less informative about the latent and intensity function than locations where the prior predictive probability for non-zero observations is significantly above zero.

\section{Survey designs}\label{sec:design}

\subsection{Spatially balanced designs and random designs} 

Our goal is to develop an algorithm that generates reasonable designs without numerically hard and time consuming optimization of the utility or cost function.
We begin the development from common random and spatially balanced designs, which are summarized in this section. 
In the next section, we introduce our extension to them to achieve better sampling designs for LGCP modeling.

We denote the uniform random sampling of design locations over the study domain by \emph{Random}. The random sampling does not typically lead to equal coverage of sampling locations, for which reason many spatially balanced designs have been introduced as alternatives to random sampling  \citep{cambanis1985,muller2007}. 
Common examples are quasi-random number sequences such as the Sobol and Halton sequences \citep{sobol1976} used in this work. We denote them by \emph{Sobol} and \emph{Halton}.
One of the most popular spatially balanced designs is the Fibonacci lattice. 
For 2-D square it is detailed by, for example, \citet{koehler1996} and \citet{pei2009} define an algorithm to construct the Fibonacci lattice in 3D setting. 
We use their algorithm here and denote the corresponding design \emph{Fibo\_lat}.

We include into comparison also two recent distance-based design methods; the simple inhibitory and the inhibitory plus close pairs designs \citep{chipeta2016}, denoted here by \emph{min\_dran} and \emph{close\_pair} respectively. 
These designs introduce a minimum dispersion threshold in the random sampling of design locations. 
For example, under the simple inhibitory design, the distance between any two locations should be greater than or equal to the threshold. 
\citet{chipeta2016} showed that the designs generated by these methods have good performance in parameter estimation and spatial prediction with Gaussian observation model.
Their algorithm was tailored for continuous covariate space so we extended it to work also for discretized locations
to be denoted by \emph{min\_dist}. 
The distance threshold and number of close pairs in these designs could be optimized \cite{chipeta2015}. 
However, we fixed them based on preliminary test runs.
All distance-based designs were constructed in unit cube and then scaled to the actual size of the domain. 
The distance threshold in the cube was $\delta=0.21$ for design size $n=50$, $\delta=0.15$ for $n=100$ and $\delta=0.1$ for $n=150$. For the close\_pair design we set the number of close pair points, $k$, to $0.5\times n$ and the distance threshold to $\delta_k = \delta * \sqrt{n/(n-k)}$. 

The above designs are based on either random or quasi random sequences and can easily be used as proposal algorithms in rejection sampling. 
We add into comparison also one deterministic space-filling design due their popularity. Space-filling designs is a class of purely geometrical designs using distance-based criteria to search for uniform spatial coverage \citep{royle1998,nychka1998,muller2007,johnson1990}. 
\citet{muller2001} and \citet{muller2007} summarize these designs by a numerical search algorithm called \lq \lq Coffee-house\rq \rq ~which is used in this work and called \emph{space\_fill}. 


\subsection{Rejection sampling designs}\label{sec:reject}

Balanced designs perform well in maintaining the spatial regularity. This may, however, be suboptimal if some locations are expected to be more informative than others.
In order to account for the specific properties of the expected utility under the LGCP model (Section~\ref{sec:properties_of_utilities}) we extend them so that on average more survey sites are located to places which are expected to increase the utility the most.
In practice, we extend the idea of balanced acceptance sampling \citep{robertson2013a,robertson2013b} and propose a new and more general design method called \lq \lq rejection design\rq \rq, ~where a spatially balanced design is thinned with an inclusion probability that is a function of the prior mean of the latent function or intensity function in LGCP. 

The general algorithm of the rejection sampling design proceeds as following:
 \begin{enumerate}
 \item Randomly generate a location $\x^{\ast}$ within the study domain (here any of the above random or quasi-random sequence can be used); 
 \item Calculate an inclusion probability $0\leq p(\x^{\ast})\leq 1$
 \item Accept the location with probability $p(\x^{\ast})$. If accepted, set $\x_j=\x^{\ast}$ and increase $j=j+1$. If rejected, keep $j=j$ and return to step 1;
  \item Repeat steps 1-3 until the size of design reaches to $n$.
\end{enumerate}
A rejection sampling design corresponds to a thinned version of its underlying random or spatially balanced design.
The inclusion probability can be linked with prior knowledge of the intensity function and its choice governs how much weight is assigned to sample higher intensity areas. The above algorithm cannot be directly used with the deterministic space-filling design for which reason we developed a modified coffee-house algorithm for rejection sampling as detailed in Appendix~\ref{sec:space-fil-rejection}.
For the 3-D Fibonacci lattice design, we used dynamic scaling \citep[][]{family1985} to obtain a design with inclusion probability restricted to unit cube.

We tested three inclusion probability functions: an inclusion probability proportional to the expectation of the latent function 
$p(\x)\propto\mu(\x)$, 
an inclusion probability proportional to the expected intensity, 
$p(\x)\propto e^{\mu(x)+2\sigma^2(x)} $, and an inclusion probability proportional to truncated expected intensity 
$p(\x)\propto\min\left(p_{\mathrm{max}},e^{\mu(x)+2\sigma^2(x)}\right)$
 where $\mu(\x)$ and $\sigma^2(x)$ are either the prior  (when there is no earlier data) or posterior (when old data exist) mean and variance of the GP and $p_{\max}$ is a tuning parameter.
If $\mu(\x)$ is negative at some $\x$, proper scaling on $\mu(\x)$ is necessary to keep $\{\mu(\x)\geq 0, ~\forall \x \in \D\}$ in the first inclusion probability function.
Each of these inclusion probabilities give more weight to locations with higher $\E[Y]$ which should provide more informative data as discussed in Section~ \ref{sec:properties_of_utilities}.
The inclusion probability proportional to $\mu(x)$ weights the high intensity locations least and, hence, modifies the underlying random or balanced design the least. 
The inclusion probability proportional to expected intensity weights the high intensity locations the most.
If there are large differences in the intensity function this inclusion probability can lead to sampling designs that are too concentrated in only a small portion of the study domain. 
For this reason, we introduce an inclusion probability proportional to the truncated expected intensity. The tuning parameter $p_{\max}$ governs how much weight is given to the highest intensity locations.
For example, in one of the case studies over 90\% of the prediction domain would have inclusion probability less than 5\% if the probability was formed proportional to the expected intensity. 
By setting $p_{\max}$ we can control the portion of design points within the low intensity region which forms the majority of the study domain.

\section{Simulation studies}\label{sec:simulation_studies}

\subsection{Study setting}

In this section, we study the properties of spatially balanced designs and their rejection sampling versions introduced in section \ref{sec:design} with simulation study.
The study was carried out in the unit cube $[0,1]^3$ with both the separable~\eqref{eq:separable} and the additive~\eqref{eq:add} GP prior.
We use a Mat\'ern \citep{matern2013} covariance function 
$k_{\nu=3/2}(\s,\s')=  \sigma_s^2\left(1+\frac{\sqrt{3}|\s-\s'|}{l_{\s}}\right)\exp\left(\frac{\sqrt{3}|\s-\s'|}{l_{\s}}\right)$ in the spatial domain
and a Gaussian covariance function $k_t(t,t')=   \sigma_{t}^2 e^{-(t - t')^2/l_{t}^2} $ in the temporal domain. The positive parameters $l_{\s}$ and $l_t$ are characteristic length-scales \citep{Banerjee+Carlin+Gelfand:2015}, which affect the correlation structures, and the positive parameters $\sigma_s^2$ and $\sigma_t^2$ are variance parameters that govern the magnitude of process variations. 
In the separable model we set $\sigma_s^2 =1$ for identifiability.
 
\begin{figure}[!t] 
 \centering
\includegraphics[width=0.8\textwidth]{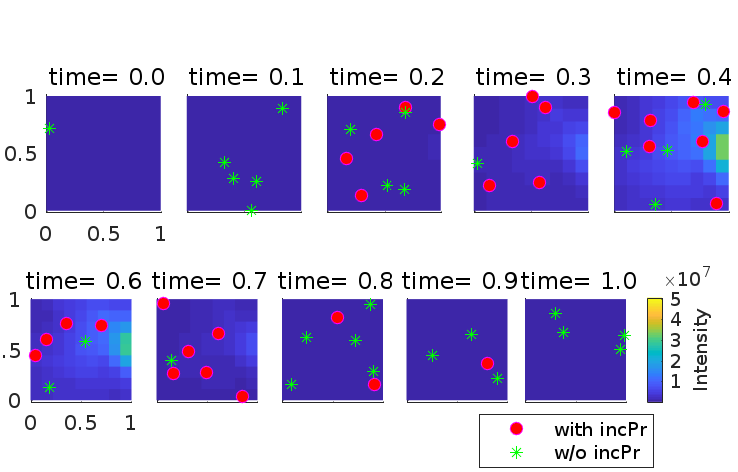}
 \caption{\label{fig:colormap} A random draw from an additive GP with unimodal mean function along time (color surface) and samples from Sobol design ($n=30$) with (red dots) and without (green asterisks) rejection sampling. 
 The inclusion probability, $p(t)\propto \mu(t)$.}
\end{figure}

We used a concave temporal mean function $\mu(\s,t)=\mu(t)=a-c(t-b)^2$ with parameters $a=2, b=0.5$ and $c=30$ so that the prior predictive probability of $y>0$ is almost zero at the start and at the end of the time period $t\in [0,1]$ but $\E[y]$ is clearly above zero in the middle of the time period. See an example in Figure \ref{fig:colormap}. 
In order to gain understanding on performance of alternative designs with different kinds of spatiotemporal random effects we tested the designs with a set of alternative covariance function parameter values. The tested temporal range parameters were $l_t=\{0.2, 0.85, 1.5\}$, temporal variances were $\sigma^2_t = \{0.5, 1, 2\}$ and spatial range parameters were $l_{\s}\in \{0.2,0.4,\dots,1.6\}$. The spatial variance was fixed at $\sigma_{\s}^2 = 2$ in all experiments. 
The inclusion probability used in the simulation studies is proportional to the expectation of the latent function; that is $p(\x)\propto\mu(t)$.
An example, of a random draw from an additive GP ($l_t=1$, $l_s=1$ and $\sigma_s^2=2$, $\sigma_t^2=1$), together with Sobol design with and without rejection sampling is presented in Figure~\ref{fig:colormap}. 
It depicts the allocation of more samples to times with high inclusion probabilities using rejection sampling.
Both designs cover the whole unit square. 

Each design was evaluated with design sizes $n=50, 100$ and $150$ using the expected APV loss \eqref{eq:u21} and the expected KL-divergence \eqref{eq:ekl1}  utility. 
In order to compare the effect of Poisson likelihood to the optimal design we evaluated the designs also with equal GP models with a Gaussian observation model.
We only show the results of $n=100$. Results with other design sizes were similar but the expected losses were smaller and expected utilities larger with increasing design size. 
We applied the GPstuff toolbox \citep{vanhatalo2013} in the calculations here and in the case study.

%

\subsection{Results, average predictive variance}

\begin{figure}[!ht] 
 \centering
\includegraphics[width=0.9\textwidth]{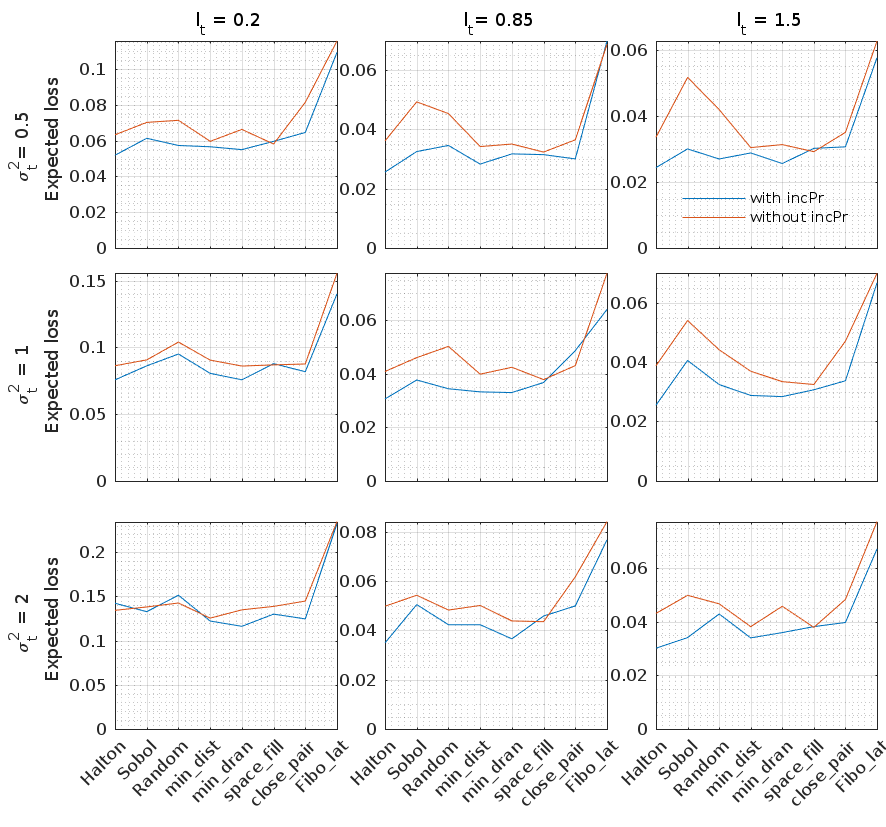}
  \caption{\label{fig:Poisson_APV_additive} The expected APV loss of Poisson intensity ($\bar{L}_{\text{APV}\lambda}$) of a model with additive covariance function at different values of $l_t$ and $\sigma_t^2$ averaged over $l_{\s}\in \{0.2,0.4,\dots,1.6\}$ when using designs with and without inclusion probability and $n=100$. }
\end{figure}

\begin{figure}[!ht] 
 \centering
\includegraphics[width=0.9\textwidth]{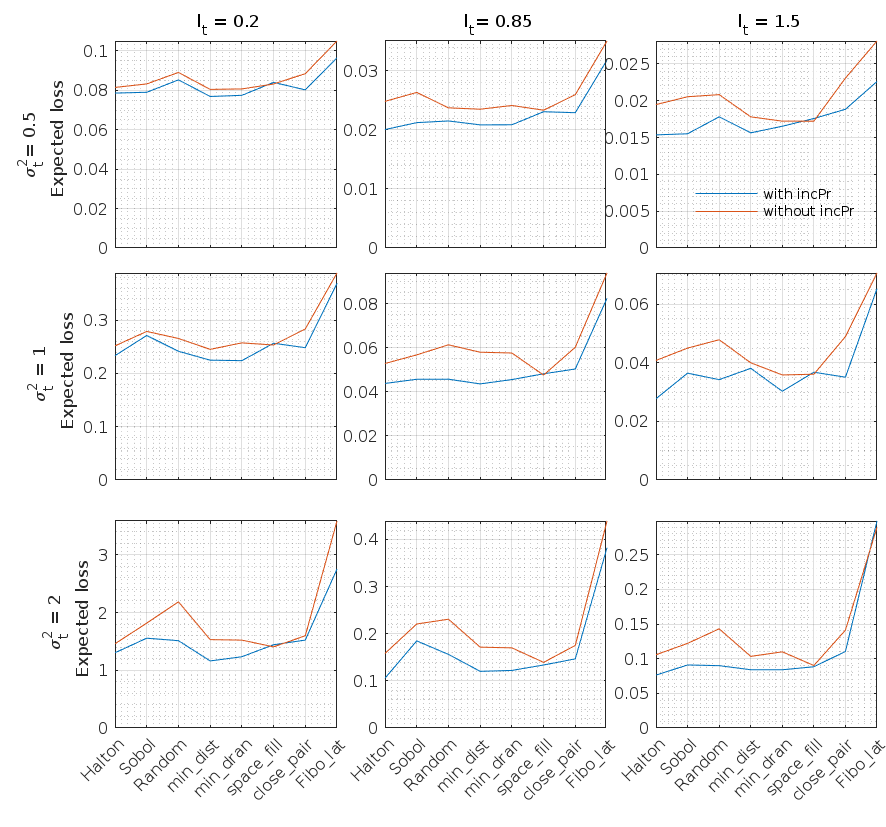}
  \caption{\label{fig:Poisson_APV_separable} 
  The expected APV of Poisson intensity ($\bar{L}_{\text{APV}\lambda}$) of a model with separable covariance function at different values of $l_t$ and $\sigma_t^2$ averaged over $l_{\s}\in \{0.2,0.4,\dots,1.6\}$ when using designs with and without inclusion probability and $n=100$. }
\end{figure}


%
The differences between designs were qualitatively similar whether considering the APV of latent function or intensity so here we show only the latter (see Supplement for former).
Figures \ref{fig:Poisson_APV_additive} and \ref{fig:Poisson_APV_separable} show the expected APV of intensity for designs with and without inclusion probability averaged over $l_{\s}\in \{0.2,0.4,\dots,1.6\}$ for different values of $l_t$ and $\sigma_t^2$. 
Results for each $l_s$ separately are given in Supplement. 
The expected APV loss decreases when using rejection sampling compared to not using rejection sampling with all other designs except the space\_fill design in which case they are equal.

The Halton and minimum distance designs (min\_dist and min\_dran) perform better than the alternatives with smaller loss, while the Fibonacci lattice design (fibo\_lat) is among the worst.
The Sobol designs are in general the second worst compared with the rest of the designs. 
The results show the decrease in expected APV loss in designs with rejection sampling compared to corresponding designs without rejection sampling ranges from nearly zero (space\_fill design) to approximately 20-30 \% (rest of the designs). 
The rejection sampling improves the efficiency of the best spatially balanced designs considerably. 

The relative difference in expected APV between designs with and without rejection sampling increases with increasing $l_t$ and decreasing $l_s$ (see also figures in Supplement) as well as with increasing $\sigma_t^2$. 
This is reasonable since when $\sigma_t^2$ is increased the prior uncertainty about intensity increases the most at times with highest prior latent mean, $\mu(t)$.
As temporal length-scale, $l_t$, increases the temporal variation of $f(\x)$ around $\mu(t)$ gets smaller and observations at times around the prior predicted peak intensity time inform more about the spatial variation around $\mu(t)$ at other times as well.
This is especially evident in the additive model where the spatial structure of the spatiotemporal random effect is the same, $g(\s)$, throughout the time interval and only its level, $h(t)$, changes. 
With increasing $l_t$ the spatiotemporal random effect approaches temporally constant spatial random effect in which case sampling at times when we expect to see most spatial variation in observations inform about the structure of spatiotemporal random effect at other times as well.
With increasing spatial length-scale $l_s$ the spatial variation in $\lambda(\x)$ decreases and the prior uncertainty about spatial structure decreases as well.

To summarize, the rejection sampling algorithm is expected to be the more beneficial the more spatial variation and the less temporal variation we expect $f(\x)$ to have around $\mu(t)$.
Similarly, if the prior mean has only spatial structure so that $\mu(\x) = \mu(\s)$ the rejection sampling with inclusion probability proportional to $\mu(\s)$ is expected to be the more beneficial the more temporal variation and the less spatial variation we expect $f(\x)$ to have around $\mu(\s)$.
When prior mean varies in both space and time, rejection sampling with inclusion probability proportional to $\mu(\x)$ is expected to decrease expected APV compared to designs without rejection sampling.
 This is illustrated in the case study experiments in Section~\ref{sec:case_study}.

\subsection{Results, KL-divergence}

\begin{figure}[!ht] 
 \centering  
\includegraphics[width=0.9\textwidth]{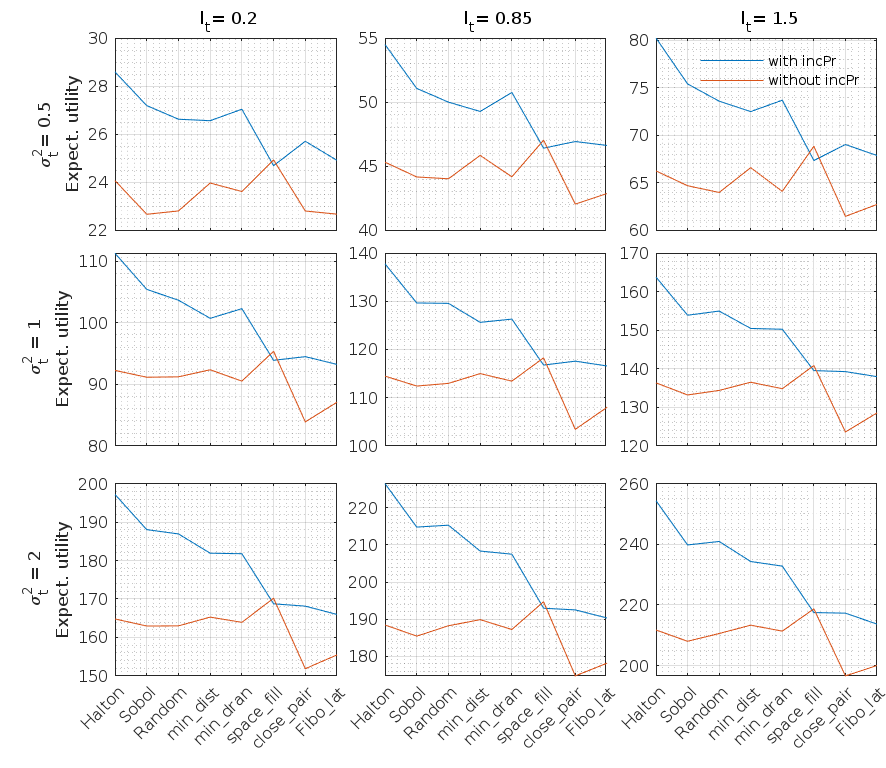}
  \caption{\label{fig:Poisson_KL_additive} The expected KL-divergence utility of intensity ($\bar{U}_{\text{KL}\lambda}$) of a model with additive covariance function at different values of $l_t$ and $\sigma_t^2$ averaged over $l_{\s}\in \{0.2,0.4,\dots,1.6\}$ when using designs with and without inclusion probability and $n=100$. }
\end{figure}

\begin{figure}[!ht] 
 \centering
\includegraphics[width=0.9\textwidth]{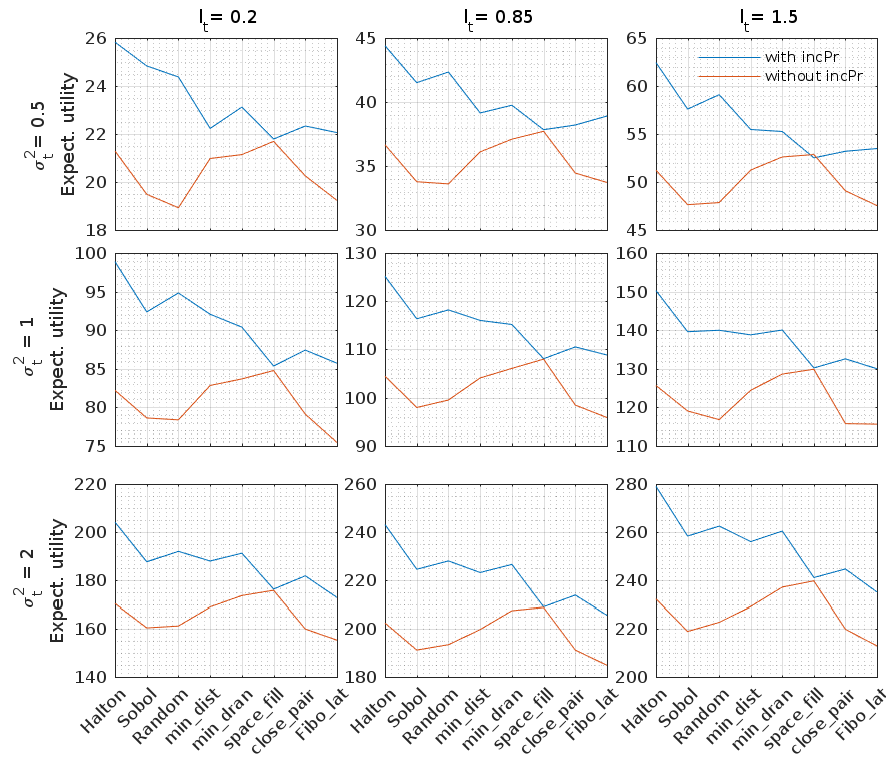}
  \caption{\label{fig:Poisson_KL_separable} The expected KL-divergence utility of intensity ($\bar{U}_{\text{KL}\lambda}$) of a model with separable covariance function at different values of $l_t$ and $\sigma_t^2$ averaged over $l_{\s}\in \{0.2,0.4,\dots,1.6\}$ when using designs with and without inclusion probability and $n=100$.}
\end{figure}

Figures~\ref{fig:Poisson_KL_additive} and~\ref{fig:Poisson_KL_separable} show the expected KL-di\-ver\-gen\-ce $\bar{U}_{\text{KL}}$ from prior to posterior under the different designs. 
The designs with rejection sampling work again better than the designs without inclusion probability. 
The expected KL-divergence is approximately 20 \% smaller in designs without rejection sampling compared to designs with rejection sampling.
The differences between alternative designs are also larger in the expected KL-divergence than in the expected APV loss. 
Halton design is again the best and Fibonacci design is the worst. 
Other well performing designs are Sobol, Random and minimum distance (min\_dist and min\_dran) designs.
The best performing rejection sampling algorithms use balanced designs that produce spatially the most uniform allocation of sampling sites among the alternatives.
They also provide the most information about the posterior covariance structure of $f(\x)$.


\subsection{Comparison to models with Gaussian likelihood}
Contrary to the results concerning the Poisson likelihood, with Gaussian likelihood the expected APV loss increases and the expected KL-divergence decreases when using rejection sampling compared to not using rejection sampling (see Supplement).
This is well in line with earlier results on optimal spatial designs \citep{diggle2006} since with Gaussian likelihood, data are equally informative everywhere in the whole study domain regardless of $\mu(\x)$ and we learn the most about the latent function by sampling the domain \lq\lq uniformly\rq \rq~ (see Section~\ref{sec:properties_of_utilities}).
Hence, the optimal sampling designs can be very different under the Gaussian and LGCP models.

\section{Case study on fish reproduction areas}\label{sec:case_study}

\subsection{Case study data and model}

\begin{figure}[t]
\centering
\includegraphics[scale=0.4]{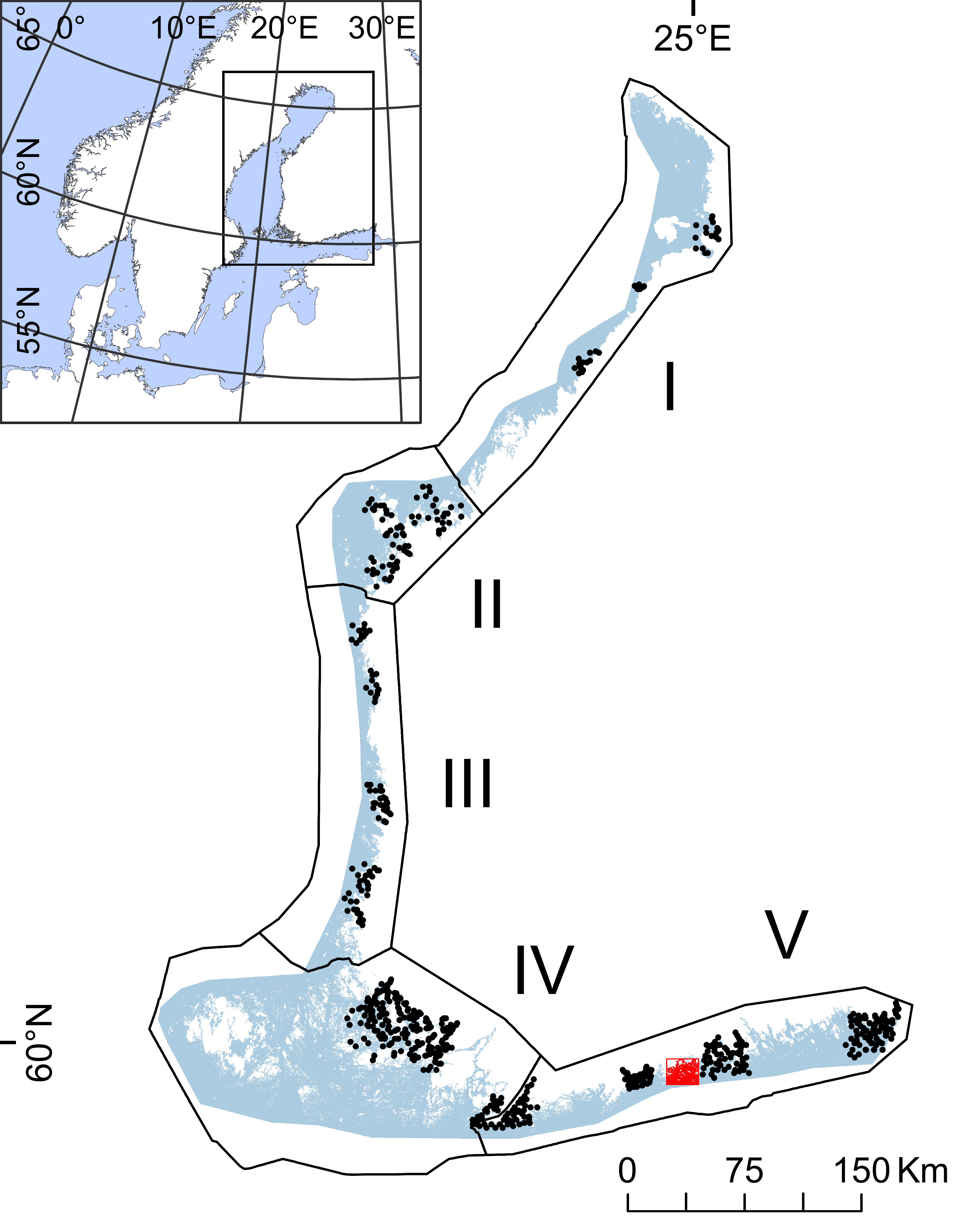}
\caption{Map of the case study area on the Finnish coastal region. The black dots show the sampling locations of existing data and the red square shows the region over which we want to plan a new sampling design. }\label{fig:case_study_area} 
\end{figure}

In the case study, we compared the above sampling designs for analyzing distribution of larval areas of pike perch (\emph{Sander lucioperca}) and Baltic herring (\emph{Clupea harengus membras}) on Finnish coastal region in the northern Baltic Sea (see Figure~\ref{fig:case_study_area}). 
Pike perch and herring are commercially important fish species and information on their larval areas is needed for sustainable fisheries management. 
\citet{Kallasvuo+etal:2017} studied the distribution of these species along the Finnish coastal region and showed that in case of pike perch, the most important reproduction areas are extremely local whereas herring reproduction areas are rather uniformly distributed.
Hence, efficient sampling designs for these two species are expected to be rather different.

The existing species distribution data (n=1788) were collected during years 2007-2014.
The locations of sampling sites vary between years and, since the exact time of larval hatching is not known, within a year each sampling site was visited several times between the calender days 128 (early May) and 188 (early July).
Each sampling site is a transect of length 400-500 meters along which a net with 0.028 m$^2$ opening was towed behind a boat. The net sampled the surface water (depth 0.5-1.0m) and the species observations consist of the number of larvae in the volume of sampled water. The sampling sites were combined with seven abiotic environmental covariates that were available as raster maps with resolution of 50 m throughout the Finnish coastal region.
Here, we extend the work of \citet{Kallasvuo+etal:2017} by planning a new sampling design to improve the distribution estimates in a region $\tilde{\A}\subset \A$ that was not included in earlier data collection.
The new sampling region is located near Helsinki and is approximately 40 km wide and 40 km long.
We want to plan a spatiotemporal sampling design within the subregion $\tilde{\A}$ (Figure~\ref{fig:case_study_area}) between calendar days 100-240 (from early April to the end of August) comprising the prediction domain $\sD = \tilde{\A}\times [100,204]$.

Following \citet{Kallasvuo+etal:2017} we modeled the observed larval counts as overdispersed Poisson process. 
Due the small size of sampling transects compared to total study region each sampling site is treated as one discretized location in \eqref{eq:likelihood}. 
At $i$'th sampling site the observed number of larvae is $Y_i\sim \text{Poisson}(V_i\lambda(\x_i) \epsilon_i)$. Here, $V_i$ is the sampled volume of water, $\lambda(\x_i)$ corresponds to the intensity of the Poisson point process at location $\x_i$ as predicted by the environmental covariates and spatiotemporal location and $\epsilon_i$ is a random effect corresponding to the $i$'th sampling occasion. 
The independent random effects describe, for example, non-structured stochasticity due to environmental conditions during the sampling. 
Since volumes $V_i$ are approximately equal we gave a joint prior for the random effects with $\epsilon_i \sim \text{Gamma}(r,1/r)$ where the Gamma distribution is parameterized with scale and shape so that $\E[\epsilon_i]=r\frac{1}{r}=1$ and $Var[\epsilon_i]=1/r$. We can now write $Y_i\sim \text{Poisson}(\tilde{\lambda}_i(\x) )$ where $\tilde{\lambda}_i(\x) \sim \text{Gamma}(r,\lambda/r)$ and marginalize over $\tilde{\lambda}_i(\x)$ to get $Y_i \sim \text{Negative-Binomial}(\lambda, r)$ where the Negative-Binomial distribution is parameterized so that $\E[Y_i]=V_i \lambda_i$ and $Var[Y_i]=\E[Y_i] + \E[Y_i]^2/r$.
Hence, $r$ is an overdispersion parameter corresponding to, for example, multiplicative independent random errors in observations \citep{Linden+Mantyniemi:2011}.
When the dispersion parameter $r \rightarrow +\infty$, the Negative Binomial approaches Poisson distribution.
The likelihood function is now
\begin{equation}
L(\y|f(\cdot), V, r) = \prod_{i=1}^n \mbox{Negative-Binomial}(V_i e^{f_i}, r).
\end{equation}

The log intensity was given a zero mean additive GP prior 
\begin{align*}&
f(\z,\s,t,\tau) \sim &\\& GP \left(0, \sigma_{\alpha}^2 + \sum_{j=1}^7k_j(z_j,z_j') + k_8(\z,\z') + k_9(\s,\s') +  k_{10}((s,t),(s',t')) +  k_{11}(\tau,\tau')\right),
\end{align*}
where $\z\in \Re^7$ is the vector of  environmental covariates, $t$ corresponds to year and $\tau$ corresponds to the day of a year. The additive components are: $\sigma_{\alpha}^2$ is the prior variance of intercept, $k_1,\dots,k_7$ are Gaussian covariance functions related to additive covariate effects, $k_8$ is Gaussian covariance function of joint covariate effect, $k_9$ is a M{\'a}tern, $\nu=3/2$, covariance function of spatial random effect, $k_{10}$ is a separable spatiotemporal covariance function formed by M{\'a}tern, $\nu=3/2$, (spatial) and exponential (temporal) functions, and $k_{11}$ is a Gaussian covariance function for the effect of a day within a year. \citet{Kallasvuo+etal:2017} modeled larval distribution only during their (approximate) peak abundance and did not include the last additive term. 
It was included here in order to model the development of larval abundance within a year which then provides information when the future sampling should be done. 
We gave weakly informative priors for the covariance function parameters so that inverse of length-scales and variance parameters were given half Student$_{\nu=4}$-$t$($\mu=0,s^2=1$) prior distributions.

The sampling days in the existing data are distributed rather sparsely from early May to the end of July. 
A regular GP prior for the calendar day effect might give ecologically unreasonable results due to the property of radial basis covariance functions to revert the GP prediction to the the prior mean far from data. 
For this reason, we imposed a functional constraint for the calendar day effect that forces it to have positive (negative) derivative at the beginning (the end) of the potential sampling period.
The joint distribution of the latent function and its derivative $d f(\z,\s,t,\tau)/d \tau$ is a Gaussian process \citep{rasmussen2006} so we can impose the monotonicity constraint by using virtual derivative observations \citep{riihimaki2010,shively2009}
We set in total 10 virtual observations for pike perch every ten days between calender days $[100, 130]$ and $[190, 240]$, whereas for Baltic herring we use 7 virtual observations between days $[100, 120]$ and $[210, 240]$.
At the virtual observation locations within the former limits the derivative of the latent function was given a probit likelihood $\Phi(\rho^{-1}d f/d \tau)$ and within the latter limits the derivative was given a likelihood $\Phi(-\rho^{-1}d f/d \tau)$. 
The scaling parameter $\rho$ governs how closely the standard Gaussian cumulative distribution function ($\Phi(\cdot)$) approximates the step function  and it was set to $\rho=10^{-6}$ \citep{riihimaki2010}.

\subsection{Methods in the case study}

Since we have old data to inform about the intensity function we base the choice of the design on posterior instead of prior predictive utility/loss. 
The prior predictive distribution $p(Y|d_n)$ in equations \eqref{eq:APVloss} and \eqref{eq:KL2} was replaced with the posterior predictive distribution $p(Y|\mathbf{y},d_n)$, where $\mathbf{y}$ denotes the existing data. Similarly, $Var(f|d_n,Y)$ and $p(f|Y,d_n)$ were replaced by $Var(f|d_n,Y,\mathbf{y})$ and $p(f|Y,d_n,\mathbf{y})$. 
In the rejection sampling designs, we tested all three inclusion probabilities introduced in Section~\ref{sec:reject}.
The rejection sampling worked better than the corresponding balanced design without rejection sampling in each case. We report the results only for the best inclusion probability that was proportional to truncated expected intensity
\begin{equation}\label{eq:caseStudy_inclusionProb}
p(\z,\s,\tau) \propto \min\left(p_{\max},e^{\E[f(\z,\s,\tau)|\mathbf{y}]+2\mathrm{Var}\E[f(\z,\s,\tau)|\mathbf{y}]}\right). 
\end{equation}
The truncation threshold $p_{\max}$ was set to 0.15 for pike perch and to 0.5 for herring.
We used the Laplace method to form approximation for the posterior of the latent function so that we optimized the hyperparameters to their (approximate) MAP estimate and conditional on this estimate approximated the posterior of latent function \citep[][]{Vanhatalo+Pietilainen+Vehtari:2010}. 

We compared the alternative spatially balanced designs with and without rejection sampling using the expected APV loss and the expected KL-divergence with $n= 100$ design points.
When constructing designs, we scaled the design space to the unit cube and used the same distance thresholds as in the simulation study. 
The prediction domain is not continuous and includes land areas that need to be ruled out (Figure~\ref{fig:monotonic1}).
First we used the balanced sampling design methods to produce candidate points from a cube that covers the subdomain $\sD$ and the time interval of interest. 
We then applied the Branch-and-Prune method \citep{kubica2014} to rule out the land areas (and the sea area out-of scope of the study, see Figure~\ref{fig:cased4}). 
In rejection sampling, the reject rule was then applied on each candidate point left.
These steps were continued so long that we had as many design points as wanted. 

The expected APV loss was calculated similarly as in the simulation studies. With temporal resolution of one week and spatial resolution of 50 meters the total number of 3D grid cells was 4\,588\,580.
Since the prediction domain $\sD$ does not include all data points (the old data falls outsize the study region), we would need to use \eqref{eq:klfield} to calculate it instead of \eqref{eq:KL} that was used in the simulation studies. 
For this reason we would need to approximate also the KL-divergence on the 3D grid.
Due to the size of the grid the required covariance matrix inversion was infeasible for which reason we report results only for APV loss. 

\subsection{Results}


\begin{figure}[t] 
\begin{minipage}[b]{1\linewidth}
 \centering
\includegraphics[width=0.7\textwidth]{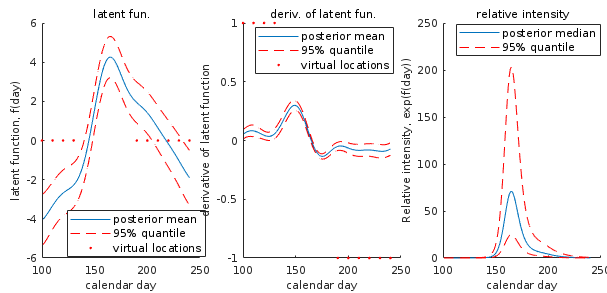}
\centerline{(a) The calendar day effect with monotonicity information.}\medskip
\end{minipage}
 \begin{minipage}[b]{1\linewidth}
 \centering
\includegraphics[width=0.7\textwidth]{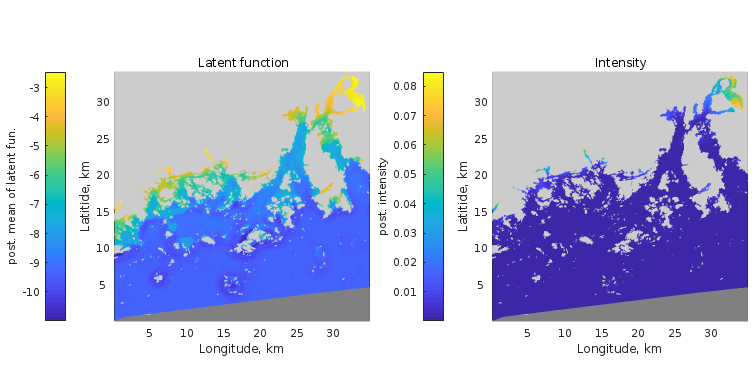}
\centerline{(b) The latent and intensity function at peak larval week.}\medskip 
\end{minipage}
 \caption{\label{fig:monotonic1} Subplot a) shows the calendar day effect on larval abundance of pike perch, its derivative, and the corresponding relative intensity changes in larval abundance. The plots show also the virtual observation locations used to code the monotonicity information. 
 Subplot b) shows the expected posterior mean of the latent function and its corresponding intensity in the prediction region $\tilde{\A}$ on calendar day 165. }
\end{figure}

\begin{figure}[t]  
 \centering
\includegraphics[width=0.7\textwidth]{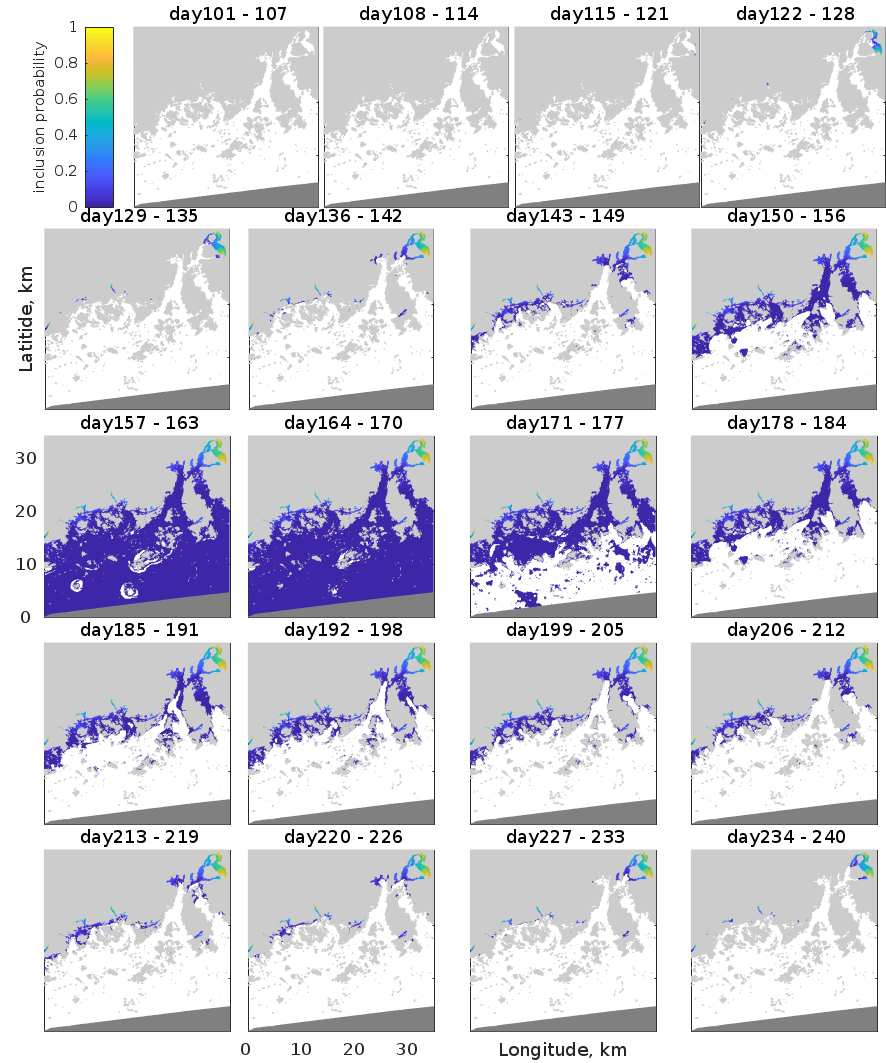}
\caption{\label{fig:case11} 
The weekly inclusion probabilities \eqref{eq:caseStudy_inclusionProb} for rejection sampling design. 
The light gray areas are land, the dark gray color indicates sea area out of scope of this
study and the white color shows the sea areas where inclusion probability is less than 0.05\%.}
\end{figure}

Figure~\ref{fig:monotonic1} summarizes the posterior of the calendar day effect and the intensity function across the prediction region on the peak larval season of pike perch.  
Figure~\ref{fig:case11} shows the weekly inclusion probability surfaces for the rejection sampling design in case of pike perch (the corresponding figures for Baltic herring are in Supplement).
There are clear spatial and temporal differences in the intensity function that are transferred to inclusion probability.
Due to strong variation in larval density, the inclusion probability is significantly above zero only during and near the peak larval period and decreases to practically zero ($<$0.05\%) over most of the region in the beginning and in the end of the study period.

\begin{figure}[t] 
\begin{minipage}[b]{0.48\linewidth}
 \centering
\includegraphics[width=1\textwidth]{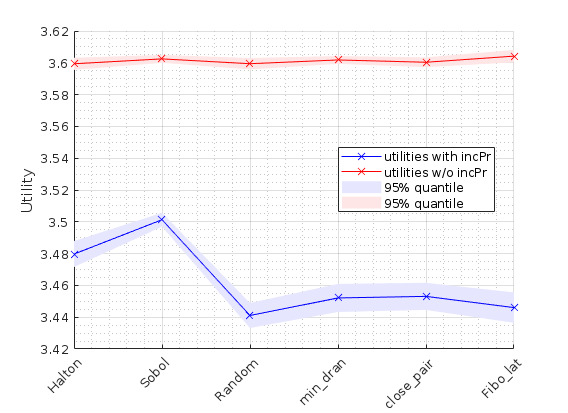}
\centerline{a) pike perch comparison}\medskip
\end{minipage}
\begin{minipage}[b]{0.48\linewidth}
 \centering
\includegraphics[width=1\textwidth]{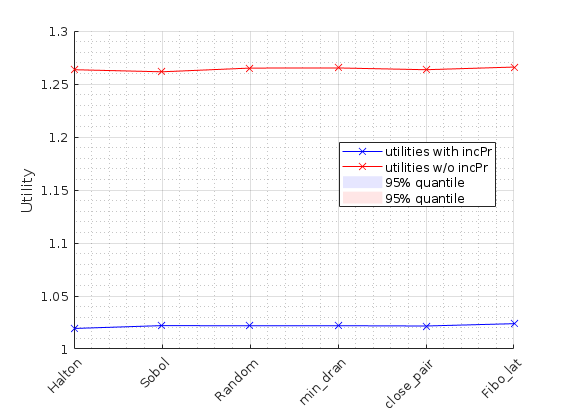}
\centerline{b) herring comparison}\medskip
\end{minipage}
 \caption{\label{fig:uKUHA}The APV loss of latent function for alternative designs in case of pike perch (a) and Baltic herring (b). The crosses connected with solid lines show the Monte Carlo estimate and the highlighted regions show the 95\% credible interval of this estimate.  }
\end{figure}

The expected APV losses of the latent function are shown in Figure~\ref{fig:uKUHA}. 
The results for APV loss of the intensity function were qualitatively similar so we omitted them here.
In general, the designs with rejection sampling work best in the case study as well.
However, there are clear differences in the performance of alternative designs. Contrary to simulation studies and herring sampling, Halton and random Sobol designs are not expected to be as good as other designs for pike perch. 

\begin{figure}[t]
\centering
\includegraphics[width=0.7\textwidth]{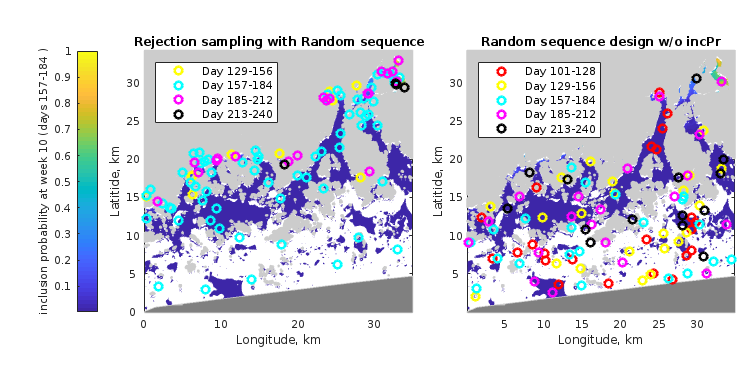}\\
\centerline{a) pike perch sampling design}\medskip
\includegraphics[width=0.7\textwidth]{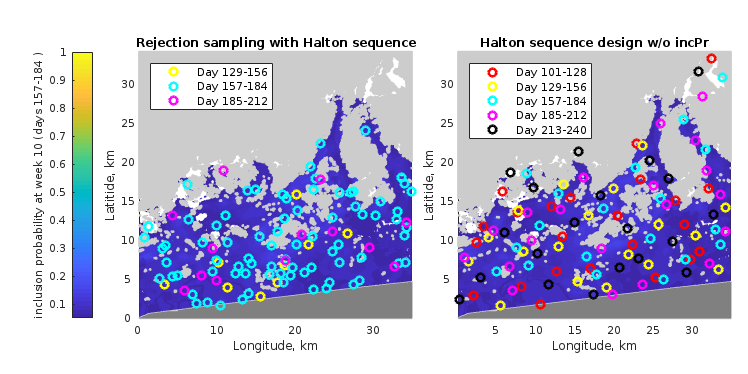}\\
\centerline{b) herring sampling design}\medskip
\caption{\label{fig:cased4}The spatial configuration of design points with and without rejection sampling for each week during the sampling period. 
Light gray areas are land and dark gray is sea area out of the prediction region. The water areas are colored according to the inclusion probability at day 165/171 with white corresponding to less than 0.1\% / 5\% inclusion probability for pike perch and herring, respectively. }
\end{figure}

Figure~\ref{fig:cased4} shows the spatiotemporal configuration of the optimal rejection designs and the corresponding balanced designs without rejection sampling for pike perch (Random) and Baltic herring (Halton). 
Table~\ref{tab:design} summarizes the weekly distribution of sampling points for these same designs.
In the rejection sampling designs the sampling is clearly concentrated on the weeks around the peak larval period (Figure~\ref{fig:monotonic1} and Supplement).
The sampling covers the whole spatial study area only on predicted peak larval period whereas on other weeks the sampling concentrates on locations with expected high larval intensity (Figure~\ref{fig:monotonic1}). 
This is reasonable since the high larval intensity spatial locations are the most informative on calendar day effect and the peak larval period is expected to be the most informative on the spatial distribution of larvae.
The sampling locations are more evenly distributed throughout the prediction domain for herring sampling than for pike perch sampling.
The even distribution is due to less variability of the intensity function for Baltic herring with a less peaked calendar day effect than the pike perch function.

\begin{table}[t]
    \caption{\label{tab:design}Weekly distribution of the design points (number of sampling locations) for pike perch (Random design) and herring (Halton design) sampling.}
 \centering
    \begin{tabular}{ | c | c| c | c | c | p{6cm} |}
    \hline
    Calender Day & With rejection sampling & Without rejection sampling \\ 
     & (pike perch /herring) & (pike perch /herring)    \\ \hline         
    101-128  &0 / 0 & 23 / 23      \\ \hline 
    129-156 & 11 / 9 &   22 / 18     \\\hline
    157-184  & 65 / 78 &  21 /19   \\  \hline    
    185-212 & 19 / 13 &  20  / 20 \\ \hline
    213-240 & 4 / 0&  14  / 20  \\  \hline      
    \end{tabular}      
\end{table}

\section{Discussion and conclusion}\label{sec:discussion}

The LGCP is a widely used point process model in many practical applications. 
In particular, it has gained increasing interest in ecology since it is an efficient and theoretically valid approach to build species distribution models ~\citep{Warton2010,Renner2013,Simpson2015a,Yuan2016,Vanhatalo+Hosack+Sweatman:2017,Makinen+Vanhatalo:2018}.
In these applications, collecting data is typically time consuming and expensive. 
For example, the small scale sampling for 100 new data points in our case study would cost approximately 50~000 euros and the costs of larger scale applications, such as the distance sampling for whale counting \citep{Yuan2016} are easily in millions of euros. 
Hence, there is a real need for efficient sampling designs in species distribution studies. For this reason, research on sampling designs has been active in recent years \citep{Foster2017b,Williams2018,Reich+etal:2018}.
There are no previous works on model-based sampling designs for LGCPs though, and most of the existing data used in LGCP analysis are based on the classical balanced or stratified sampling designs which are optimized for (linear) Gaussian models.

We have shown that in the presence of prior information on intensity function, sampling designs that are expected to be most informative for LGCPs are different from traditional designs used for Gaussian models. 
This highlights that classical spatial designs can be inefficient for LGCPs. 
The difference is caused by larger spread of the Poisson distribution with increasing intensity.
The closer to zero the intensity is the less uncertainty there is on outcome of the future data and the less information new data is expected to provide.

For this reason, we proposed a new sampling method, a spatially balanced design with rejection sampling, which gives more weight to prior predictive high intensity areas than low intensity areas. 
Our extensive simulation and case study experiments showed that when analyzed with APV loss and KL-divergence utility the rejection sampling designs consistently outperformed the corresponding balanced designs.
The relative performance of the rejection sampling designs versus balanced designs without rejection sampling was not sensitive to the variance and length-scale of the spatiotemporal Gaussian process. With all tested combinations of fixed length-scale and variance the rejection sampling design performed better than the corresponding balanced design.
The benefits from rejection sampling were increased for larger length-scales and variances. 
The inclusion probability in our new design algorithm is based on the prior (or current posterior) mean of the intensity or its logarithm. 
The inclusion probability function can also be formulated differently but we leave more thorough studies on this for future.

Our experiments considered planning the sampling design when inferential interest is in the latent and the intensity function so all analyses were performed with fixed hyperparameters. 
In simulation studies, these were chosen from a set of alternative values and in the case study we fixed them to their MAP estimate conditional on current data.
However, optimally we should conduct full posterior analysis also for hyperparameters. 
This would be especially important if we were also interested in posterior for the hyperparameters.
One obvious future research direction is thus to study which designs are best for inferring the hyperparameters of LGCPs.

The case study has direct relevance to pike perch and herring fisheries management. The rejection sampling method introduced here is straightforward to implement and, hence, can easily be applied in other regions as well. The new data can then be used to revise species distribution maps that are used in regional marine spatial planning and coastal land use management.
The results illustrated that the rejection sampling algorithm can produce very different sampling designs for different species. 
The sampling locations for herring were more uniformly distributed than the sampling locations for pike perch. 
If these two species were to be sampled at the same time, a good joint design should be a compromise between these two designs. 
In our application, reaching the sampling locations was not an issue but in larger sampling domains the design should take into account also logistic and financial constraints. 
In theory these could be included naturally into the Bayesian model-based design planning by redefining the utility and loss functions to account for the sampling costs -- or equivalently by defining a constraint functions for designs. 
In this case, we can define an inclusion probability function that leads to rejection sampling with more probable sampling locations from areas where the data are expected to increase the utility the most.
However, in these cases, a more natural approach would be to formally optimize the design.

Model-based designs can be formulated as optimal designs where the sampling locations are chosen to maximize expected utility (or minimize expected loss) \citep{muller1999,muller2001}. 
Optimal design problems are usually computationally difficult and time consuming. 
This is because the dimension of the search space ($3^n$ in our case) is so large that finding the mode of the expected utility surface is often infeasible \citep{Reich+etal:2018}.
In most cases, the posterior distribution $p(f(\cdot)|d_n, y)$ does not have a standard form, and as a consequence, numerical approximation or stochastic sampling  algorithms in the calculation of utilities are usually required. 
Our goal was to develop a balanced design algorithm offering improvements over existing algorithms and generating reasonable designs.
The proposed design method is straightforward to implement with any LGCP model and improves the existing balanced designs. 
Our results also highlight the need for more work on spatial and spatiotemporal designs for LGCP models.

\appendix

\section{Results on KL-divergence}\label{apt:D}

\paragraph{Proof of Lemma 1} \label{apt:subc1}
Assume $f(\cdot): \D\rightarrow \Re$ is a latent function with a Gaussian process prior and denote the prior probability measure of $f(\cdot)$ in the domain $\D$ by $\p(f(\cdot))$. 
Denote by $\f =[f(\x_1),f(\x_2), \cdots, f(\x_n)]$ a vector of latent variables, at finite number of locations $d_n=[\x_1^T,\dots,\x_n^T]$ where $\x_i \in \D$, and by $Y=[Y_1,\dots,Y_n]$ a random vector of observations at these locations $d_n$.
The observations are assumed to be conditionally independent given the latent variables; that is $p(Y=y|f(\cdot))=p(Y=y|\f)=\displaystyle \prod_{i=1}^n p(y_i|f(\x_i))$, where $y=[y_1,\dots,y_n]$ is a realization of $Y$. 
Denote by $ \p(f(\cdot)| d_n,Y)$ the posterior probability measure. By Bayes theorem $ \frac{d \p(f(\cdot)| d_n,Y)}{d\p(f(\cdot))} = \frac{p(y|f(\cdot))}{p(y)}$. 
Since the posterior probability measure is absolutely continuous with respect to the prior probability measure \citep{Schervish:1995}, we can calculate the KL-divergence from the prior to the posterior by
\begin{align}&
\mbox{KL}\biggl(d \p(f(\cdot)|d_n,y)||d \p(f(\cdot))\biggr) = \int \log \frac{d\p(f(\cdot)|d_n,y)}{d\p(\cdot)} d \p (f(\cdot)| d_n,y)\notag\\ 
&=\int \log \frac{p(y|f(\cdot)) d \p (f(\cdot)) } { d \p (f(\cdot))  \int p(y| f(\cdot)) d \p (f(\cdot))} d \p (f(\cdot)| d_n,y)  &\notag\\&
= \int \log p(y| f(\cdot) ) d \p (f(\cdot)| d_n,y) - \log p(y) \notag\\
&= \int \log p(y| \f ) d \p (\f| y) - \log p(y) \label{KL-divergence-over-D},
\end{align}
where  $p(y) = \int p(y|f(\cdot))d\p(f(\cdot) = \int p(y|\f)p(\f)d\f$
 The last equality holds because $d_n \subset \D$. 
In case of Gaussian observation model $p(y_i|f_i)\sim N(f_i,\sigma^2)$, this simplifies to KL divergence between two multivariate Gaussian distributions
\begin{equation}
\mbox{KL}\biggl(d \p(f(\cdot)|d_n,y)||d \p(f(\cdot))\biggr) =\frac{1}{2} \biggl[\log |K K_1^{-1}| + \mbox{Tr} (K_1 K^{-1}) + {\bf \mu}_1^T K^{-1} {\bf \mu}_1 - n \biggr],\notag
\end{equation}
where ${\bf \mu}_1=K(K + \sigma_n^2 I )^{-1}y$ and  $K_1=K-K(K + \sigma_n^2 I )^{-1}K$  are the posterior mean and covariance of $\f$.

\begin{figure}[t]  
 \centering
\includegraphics[width=0.6\textwidth]{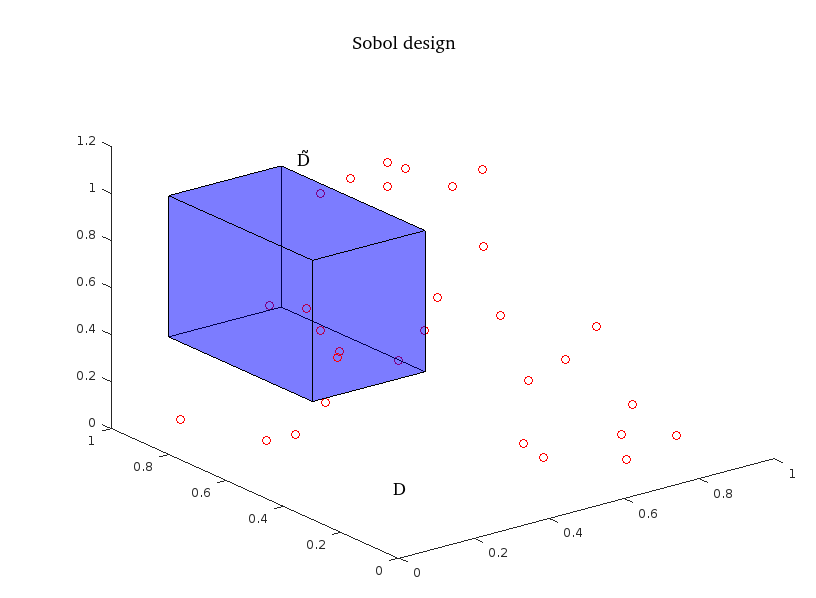}
 \caption{\label{fig:apt1}$d_n$ includes 30 locations in a unit cube ($\D$). We are interested in a region $\tilde{\D}$ marked by a blue cube which is a subset of $\D$.}
\end{figure}

Let's next consider the KL divergence from the prior to posterior of $f(\cdot)$ over a region (or subset) of locations of $\D$ that does not contain all the observations. This is illustrated in Figure \ref{fig:apt1}. We denote by $\sD\subset \D$ this subregion and by $f_{\sD}(\cdot):\sD\rightarrow \Re$ the latent function restricted to this subregion.
The KL divergence for $f_{\sD}$ is
\begin{align}&\label{eq:klfield_appendix}
\mbox{KL}\biggl(d \p(f_{\sD}(\cdot)|d_n,y)||d \p(f_{\sD}(\cdot))\biggr) 
= \int \log p(y| f_{\sD}) d \p (f_{\sD}|d_n,y) - \log p_{\sD}(y),\notag
\end{align}
where $p_{\sD}(y)= \int p(y| f_{\sD}) d\p(f_{\sD}(\cdot))$. Note that $p(y| f_{\sD})= \int p(y| \f) d \p(\f| f_{\sD})$.
The challenging part of this equation is the conditional probability measure $d \p(\f| f_{\sD})$. In practice we need to discretize the subdomain $\tilde{D}$ with fine grid cells indexed by $x_{\ast_j}$, and approximate the conditional measure by a conditional distribution $p(\f| \f_{\sD})$ where $\f_{\sD} = \{\f(x_{\ast,1}),\f(x_{\ast,2}), \cdots, \f(x_{\ast,N}) \}$. 
Hence all the above representation can be approximated numerically but with large $\sD$ and fine discretization the calculations might become infeasible. 


\paragraph{The KL divergence of intensity}\label{apt:C}
Let $\p (\lambda(\cdot))$ and $\p (\lambda(\cdot)|d_n,y)$ denote the prior and posterior probability measure of the intensity function.
As shown in Appendix~\ref{apt:D}, the KL-divergence from the prior to the posterior of the intensity is
 
\begin{align}
 \mbox{KL}(\p \left(\lambda(\cdot)|d_n,y)||\p (\lambda(\cdot))\right)
 =& \int \log p(y|\lambda) \frac{p(y|\lambda)p(\lambda)}{p(y)}d\lambda - \log p(y)\notag\\
 =& \int \log p(y|e^{f_1},\dots,e^{f_n}) \frac{p(y|e^{f_1},\dots,e^{f_n})p_f(\log\lambda)}{p(y)\prod_{i=1}^n\lambda_i}d\lambda \notag \\ &- \log p(y)\notag \\
 =& \int \log p(y|\f) p(\f|d_n,y)d\f - \log p(y),
\end{align}
where $\lambda = [e^{f_1},\dots,e^{f_n}]$ and by change of variables $p(\lambda) = p_f(\log(\lambda))/\prod_{i=1}^n \lambda_i$ and $df_i = d\lambda_i /\lambda_i $. Since  $p(y|e^x)$ and $p(y|x)$ have same $\sigma$ algebra, we have that $p(y|e^{f_1},\dots,e^{f_n}) = p(y|\f)$.

\section{Space filling rejection sampling design}\label{sec:space-fil-rejection}

The space filling rejection sampling design on discrete space is generated as follows:
\begin{enumerate}[start=0]
 \item Generate a set of candidate design locations $C=\{\tilde{\x}_j:\tilde{\x}_j\in \D\}$; for example a dense grid or a Halton/Sobol sequence.  
 \item Pick up a location $\x_{1}\in C$ from a corner of the domain and include that into the design  $d_1=\{\x_1\}$. Set $k=1$ and $i=1$.
 \item Search the location $\x_{i+1} = \displaystyle \arg \max_{\x^{\ast} \in C }^k \min_{\x_j \in d_i}   \lVert \x_j - \x_\ast \rVert$ where $\displaystyle \arg \max_{\x^{\ast} \in C }^k$ denotes the $k$'th largest value.  
\item Apply rejection sampling for $x_{i+1}$. If $ \x_{i+1}$ is rejected, set $k=k+1$ and return to Step 2. Otherwise include $\x_{i+1}$ in to the design $d_{i+1} = d_i \cup \{ \x_{i+1} \}$ and set $i=i+1$ and $k=1$; 
\item Repeat Steps 2 and 3 until the $n$'th location has been found.
\end{enumerate}

\section*{Acknowledgement}

This research was funded by Academy of Finland (Grant 304531) and the research funds of University of Helsinki (decision No. 465/51/2014). The authors acknowledge CSC – IT Center for Science, Finland,
for computational resources.


\bibliographystyle{imsart-nameyear}
\bibliography{mybibfile}  

\end{document}